\documentclass[floats,aps,showpacs,amsmath,amssymb,prl,twocolumn]{revtex4}
\usepackage{graphicx}
\usepackage{color}
\usepackage{amsmath}
\usepackage{rotating}
\usepackage{dcolumn}
\usepackage{bm}

\input epsf
\def\k{{\bf k}}
\begin{document}
\newcommand{\ltwid}{\mathrel{\raise.3ex\hbox{$<$\kern-.75em\lower1ex\hbox{$\sim$}}}}
\newcommand{\gtwid}{\mathrel{\raise.3ex\hbox{$>$\kern-.75em\lower1ex\hbox{$\sim$}}}}
\newcommand{\BSCCO}{{Bi$_2$Sr$_2$CaCu$_2$O$_{8+x}$ }}

\newcommand{\hH}{{\hat{H}}}
\newcommand{\be}{{\bf e}}
\newcommand{\bk}{{\bf k}}
\newcommand{\bp}{{\bf p}}
\newcommand{\bq}{{\bf q}}
\newcommand{\br}{{\bf r}}
\newcommand{\hDelta}{{\hat{\Delta}}}

\newcommand{\hatp}{{\hat{p}}}
\newcommand{\hx}{{\hat{x}}}
\newcommand{\hy}{{\hat{y}}}
\newcommand{\calG}{{\cal{G}}}

\newcommand{\bhatp}{{\bf \hat{p}}}
\newcommand{\bdelta}{\boldsymbol{\delta}}

\newcommand{\barv}{\overline{v}}
\newcommand{\baru}{\overline{u}}
  
\newcommand{\Fdag}{{F^{\dagger}}}

\title{Dopant-modulated pair interaction in cuprate superconductors}

\author{Tamara S. Nunner, Brian M. Andersen, Ashot Melikyan, and P. J. Hirschfeld}

\affiliation{Department of Physics, University of Florida,
Gainesville, FL 32611}

\date{\today}

\begin{abstract}

Comparison of recent experimental STM data with single-impurity
and many-impurity Bogoliubov-de Gennes calculations strongly
suggests that random out-of-plane dopant atoms in cuprates
modulate the pair interaction locally. This type of disorder is
crucial to understanding the nanoscale electronic structure
inhomogeneity observed  in BSCCO-2212,
and can reproduce observed correlations between the positions  of
impurity atoms and various aspects of the local density of states
such as the gap magnitude and the height of the coherence peaks.
Our results imply that each dopant atom modulates the  pair
interaction  on a length scale of order one lattice constant.

\end{abstract}

 \maketitle


The discovery of nanoscale inhomogeneity in the cuprates has
recently generated intense interest. In particular, the spectral
gap in the local density of states (LDOS), as observed by scanning
tunneling microscopy
(STM)~\cite{cren,davisinhom1,Kapitulnik1,davisinhom2} in \BSCCO
(BSCCO), varies by a factor of two over distances of 20-30\AA. This unusual
behavior may help reveal how the cuprates evolve from the Mott
insulating state at half-filling to the superconducting state at
finite doping. The hole concentration in the CuO$_2$-planes of
BSCCO is proportional to the number of out-of plane dopant atoms,
which also   introduce disorder.
This has led to the proposition that poorly screened electrostatic potentials
of the dopant atoms generate a variation
in the local doping concentration and thus give rise to the gap
modulations observed in STM~\cite{ZWang,QHWang,balatskyLDC}.
Poor screening has also been argued to result in
 enhanced forward scattering~\cite{AbrahamsVarma}, which appears
to be compatible with photoemission~\cite{ZHS04,scalapinoSNS} and
transport measurements~\cite{nunnermuwave} in the superconducting
state of BSCCO. An alternate perspective is explored in several
works which associate inhomogeneous electronic structure with a
competing order parameter, such as
antiferromagnetism~\cite{Kivelson,Atkinson,Dagotto}. Only very
recently  has it been possible to measure correlations between the
inhomogeneities observed in STM and positions of dopant
atoms~\cite{DavisAPS}, thus providing a clue to  the relation
between disorder and doping in this compound, as well as a means
to examine the above proposals.

In this Letter, we assume that the electronic inhomogeneity
observed by STM, at least in the optimally to overdoped samples,
can be understood  within the framework of BCS theory in the
presence of disorder. We show, however, that the conventional
modeling of  disorder as a set of random potential scatterers
fails to reproduce the most prominent features of the STM
experiments described above: (i) the subgap spectra are spatially
extremely homogeneous~\cite{davisinhom2}, unless they are taken in
the immediate vicinity of a
 defect in the CuO$_2$ plane,  (ii) the coherence peaks in regions with larger
gap tend to be much broader and reduced in height, (iii) the
``coherence peak" positions  are
 symmetric about zero bias, (iv) the dopants are found to correlate 
positively with large gap regions~\cite{DavisAPS}; and (v)
charge modulations are small~\cite{DavisAPS}. We propose that
the dopant atoms modulate the
local pair potential, i.e. the local attractive coupling $g$
between electrons is spatially dependent.
In conventional superconductors, such effects are difficult to
observe because atomic-scale modulations in $g$ produce LDOS
modulations only on the scale of the coherence length $\xi_0$.
In the cuprates, however, the situation is different due to the
short coherence length.
 We demonstrate that a model in which dopant atoms modulate the pair
 interaction gives excellent agreement with respect to the above
mentioned key characteristics of the STM data. A modulated pair
interaction could arise from local lattice distortions surrounding
the dopant atoms modifying the electron-phonon coupling or
superexchange interaction in their vicinity.

{\it Model.} We consider the following mean-field Hamiltonian for a singlet $d$-wave
superconductor
\begin{equation}
\label{eq:hamiltonian} \hat{H}\!=\!\sum_{\k\sigma}\! \epsilon_{\k}
\hat{c}_{\k\sigma}^\dagger \hat{c}_{\k\sigma} +\! \sum_{i\sigma}
\! V_i\hat{c}_{i\sigma}^\dagger \hat{c}_{i\sigma} \!+\!
\sum_{\langle ij \rangle} \! \left( \Delta_{ij}
\hat{c}_{i\uparrow}^\dagger \hat{c}_{j\downarrow}^\dagger \!+\!
\mbox{H.c.} \! \right)\!,\!
\end{equation}
where $\epsilon_\k = - 2t (\cos k_x + \cos k_y) - 4t^\prime \cos
k_x \cos k_y -\mu$ and $\sum_{\langle ij \rangle}$ denotes
summation over neighboring lattice sites $i$ and $j$. In the
remainder of the paper we will set $t^\prime/t=-0.3$ and adjust $\mu$
to model the Fermi surface of BSCCO near optimal doping
(for the homogeneous system, $\mu/t$=-1.0).
In order to account for disorder in the out-of-plane dopants,
which are separated from the CuO$_2$ plane by a distance $z$, we
include an impurity potential modeled by $V_i
=V_0\exp(-r_i/\lambda)/r_i$, where $r_i$ is the distance from a
dopant atom to the lattice site $i$ in the plane. Distances are
measured in units of $\sqrt{2}a$, where $a$ is the Cu-Cu distance.
The nearest-neighbor $d$-wave order parameter $\Delta_{ij}=g_{ij}
\langle \hat{c}_{i\uparrow} \hat{c}_{j\downarrow} -
\hat{c}_{j\downarrow} \hat{c}_{i\uparrow}\rangle$ is determined
self-consistently using (\ref{eq:hamiltonian}) with
$g_{ij}=g+(V_i+V_j)/2$.
In traditional BCS theory, $g_{ij}=g$ is spatially uniform, and
$\Delta_{ij}$ is only modulated in the vicinity of potential
scatterers~\cite{HettlerHirschfeld,Shnirman}. We will argue that
this approach is unable to reproduce observations (i)-(v) outlined
above, and
that $g_{ij}$ is strongly modified near the dopant atoms.

\begin{figure}[b]
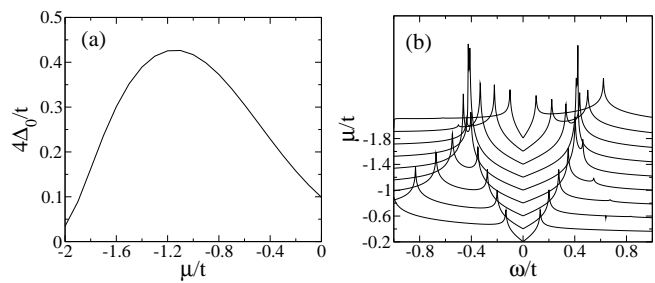

\begin{center}
\leavevmode
\begin{minipage}{.49\columnwidth}
\includegraphics[clip=true,width=.99\columnwidth]{GapMu.eps}
\end{minipage}
\begin{minipage}{.49\columnwidth}
\includegraphics[clip=true,width=.99\columnwidth]{LocalApproxMu.eps}
\end{minipage}
\caption{(a) Order parameter $\Delta_0$ for $t'/t=-0.3$ and a
constant nearest neighbor attraction of $g/t=1.16$ as a function
of chemical potential $\mu$. (b) Local density of states for
different $\mu$ using the order parameter displayed in (a).}
\label{fig:LocalApprox}
\end{center}
\end{figure}

{\it Smooth potential.} If the potential caused by the
out-of-plane dopant atoms were very smooth on the scale of
$\xi_0$, the local properties of the inhomogeneous system would be
determined by the local value of the disorder potential and the
local value of the pairing interaction. Therefore one would expect
an LDOS which is locally similar to a clean superconductor with
renormalized chemical potential $\mu-V_i$ in the case of a smooth
potential $V_i$, or with renormalized bond order parameter
$\Delta_{ij} + \delta\Delta_{ij}$ for a smooth
 off-diagonal (OD) potential.
In the case of a conventional diagonal potential, a gap size
modulation will be induced because the gap is a relatively
sensitive function  of the local chemical potential, see
Fig.~\ref{fig:LocalApprox}(a). On the other hand, modulations of this
type will inevitably have coherence peak weight-position
correlations opposite to experiment, since large gap values in the
homogeneous system imply (within BCS theory) that spectral weight
removed from low energies is transferred into the coherence peaks
(Fig.~\ref{fig:LocalApprox}(b)). This effect is further enhanced
by the presence of a van-Hove singularity at
$\omega_{vH}=\sqrt{(4t'+\mu)^2+(4 \Delta_0)^2}$ in the
tight-binding model which contributes additional weight to the
coherence peaks, in particular for $\mu/t=-1.2$ where it coincides
with the gap edge.  Here $\Delta_0$ is the bond order parameter in
the homogeneous system.  A similar although less pronounced
positive correlation between coherence peak weight and position
arises also for the smooth OD case. Note that throughout this work
we neglect inelastic scattering that would broaden the tunneling
conductance peaks at large bias but would not change their weight,
thus leaving our conclusions unaffected.

\begin{figure}[t]
\begin{center}
\leavevmode
\includegraphics[clip=true,width=1.0\columnwidth]{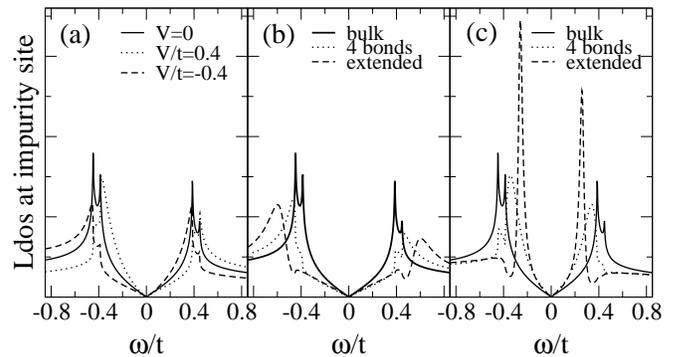}
\caption{On-site LDOS for different single-impurity models with
$t'/t=-0.3$, $\mu/t=-1$ and $\Delta_0/t=0.1$. (a) Weak pointlike
potential scatterer. (b) Dotted line: attractive ``pointlike" OD
scatterer with $\delta \Delta=\Delta_0$ on the four bonds
surrounding the impurity site. Dashed line: extended attractive OD
scatterer with $\lambda=z=1$. (c) Same as (b), with $\delta
\Delta=-\Delta_0$. } \label{fig:TMat}
\end{center}
\end{figure}

{\it Single-impurity scattering.}
Since a smooth disorder
potential cannot reproduce the experimentally observed relation
between the weight of the coherence peak
and the gap magnitude, we now address the opposite
limit, i.e. a very spiky potential 
caused by a dopant potential with short range on the scale of
$\xi_0$. Some insight into this situation can be obtained by
analyzing single-impurity scattering processes, which should be
dominant for sufficiently short ranged and weak scattering
potentials, where interference effects are negligible. For
simplicity we assume constant order parameter in the following
$T$-matrix analysis and postpone the fully self-consistent
treatment to the many impurity case.

Although solving the $T$-matrix equation for the potential
scatterer $V_i=V\delta_{i0}$ is
straightforward~\cite{BalatskyZhuVekhter}, relatively little
attention has been paid to the weak to intermediate strength
impurity case $V\lesssim t$ because it does not lead to
well-defined resonant states inside the gap.
Fig.~\ref{fig:TMat}(a) shows the LDOS at the impurity site for a
weak pointlike potential scatterer.
The positions of the coherence peaks are hardly shifted at all,
and while the spectral weight of the coherence peaks is modified,
this occurs in a distinctly particle-hole asymmetric fashion.
This is in striking contrast to the STM spectra, where
inhomogeneous but particle-hole symmetric coherence peak
modulations are observed. In addition, there is no distinct
feature in experiment corresponding to the van Hove features
present, as e.g. in Figs.~\ref{fig:LocalApprox}(b) and
\ref{fig:TMat}(a).

These shortcomings of the conventional potential
scattering model can be overcome by considering
OD scattering instead.
For the sake of clarity, in this paragraph  we neglect the
diagonal component of the potential. Fig.~\ref{fig:TMat}(b) and
Fig.~\ref{fig:TMat}(c) show the LDOS at the impurity site for a ``pointlike'' OD
scatterer with $d$-wave symmetry on the four bonds emanating from
site $i=0$, $\delta\Delta_{0,\pm\hat x} = -
\delta\Delta_{0,\pm\hat y} \equiv \delta\Delta$, and a more
extended OD scatterer with $\delta\Delta_{ij}=\pm \delta\Delta
(V_i+V_j)/(2 V_0)$, where $V_i$ is defined below
Eq.~(\ref{eq:hamiltonian}) and the negative sign applies to bonds
oriented along the $\hat y$-direction. Scattering at an order
parameter enhancement (see Fig.~\ref{fig:TMat}(b)) strongly
suppresses the coherence peaks for large values of $\delta\Delta$
or more extended OD scatterers. For scattering by a local order
parameter suppression (see Fig.~\ref{fig:TMat}(c)), exactly the
opposite happens: an Andreev resonance forms just below the gap
edge, similar to the case where the order parameter is suppressed
near surfaces~\cite{Buchholtz,Yuri}.  For large negative values of
$\delta \Delta$, or more extended OD scatterers, the Andreev
resonance moves to smaller energies, and its peak height
increases.
It draws most of its spectral weight from the van Hove singularity
at $(\pi,0)$, which is close to the part of the Fermi-surface with
the largest $d$-wave gap, i.e., the part which is most affected by
order parameter modulations. Although this indicates that the
weight of the  resonance depends on band structure, we find that
the phenomenon is very robust over a wide range of $t'$ and $\mu$.

{\it Many-impurity results.} In order to strengthen our argument
for  dopant-modulated pairing interaction we now address the
effect of self-consistency and interference between many
impurities. To this end, we solve self-consistently the
Bogoliubov-de Gennes (BdG) equations
resulting from Eq.(\ref{eq:hamiltonian}), on a $80 \times 80$
lattice rotated by $\pi/4$ compared to the Cu-O bond
direction 
(as in experimental STM maps), i.e., our system contains $2\times
80^2$ lattice sites in total. Assuming the dopants are
interstitial oxygens, each one most likely contributes two holes
to the CuO$_2$ plane. We therefore consider a dopant concentration
of 7.5\% for optimal doping which are distributed randomly in  the
reservoir layer separated from the CuO$_2$ plane by a distance
$z$.

\begin{figure}
\includegraphics[clip=true,width=0.95\columnwidth]{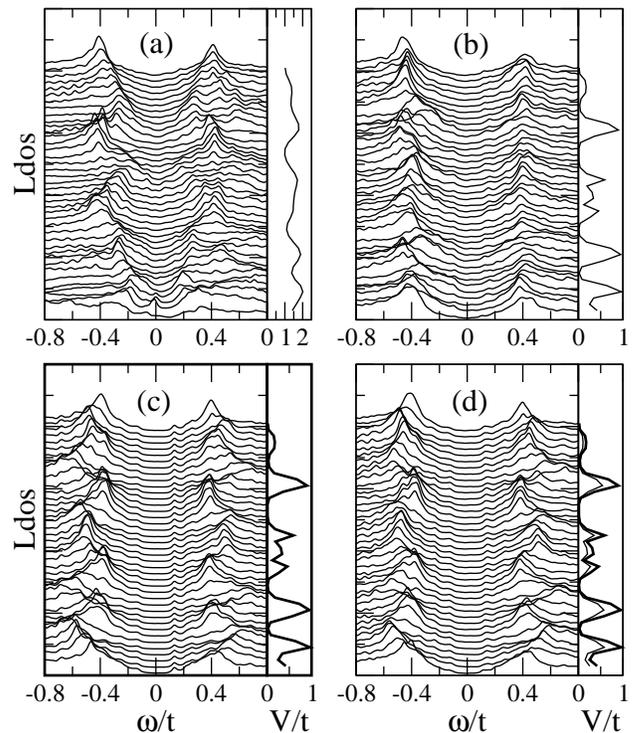}
\caption{LDOS from self-consistent solution of BdG equations,
along a straight line  for (a) conventional potential with $z=2$,
$\lambda=2$, $V_0=1.5t$; (b) same as (a), but with $z=0.57$,
$\lambda=0.5$;   (c) OD potential with $V_i$ as in (b); and (d),
combination of  OD potential shown in (c) with
 conventional potential as in (b) but with $V_0=0.6t$. Conventional (OD) potentials are
 depicted to right of each panel as a thin (thick) line.}
\label{fig:tau3example}
\end{figure}

In the limit of a smooth potential (Fig.~\ref{fig:tau3example}(a)), 
the many-impurity results agree well with the local $\mu$ picture
discussed above.
The correlation between the dopant positions and the gap amplitude
depends strongly on the size of the potential due to the
non-monotonic dependence of $\Delta_0$ on the local $\mu$, as
shown in Fig.~\ref{fig:LocalApprox}(a).
The spatial variation of the gap, however, is not rapid enough to
reproduce the grainy gap maps seen experimentally with gap
``patches" of typical size 20-30\AA~\cite{davisinhom2}; one is
therefore forced to consider ``spikier" potentials
(Fig.~\ref{fig:tau3example}(b)). In the weak limit $V\lesssim t$,
one recovers the results of the single-impurity case, i.e., the
coherence peaks are modulated in a particle-hole asymmetric way.
For the stronger spiky potentials  required  to reproduce the
magnitude of the gap modulations observed in STM, subgap states
start to form in contradiction with experiment (see
\ref{fig:tau3example} (b)).
Further discrepancies between Figs.~\ref{fig:tau3example}(a) and
\ref{fig:tau3example}(b) and the experimental spectra are: i) the
LDOS clearly does not exhibit the inverse relation between gap
size and coherence peak height; ii) the spectra are quite
particle-hole asymmetric (see Fig.~\ref{fig:tau3example}(b) and high
energy regions in Fig.~\ref{fig:tau3example}(a)); and iii) the
sizable potential required to induce gap modulations inevitably
leads to large (${\cal O}(50\%-100\%)$) local charge modulations.
The latter point puts strong constraints on any potential
scattering model, since the primary role of the impurity potential
is to couple to the density.

\begin{figure}
\begin{minipage}{.49\columnwidth}
(a)\\[-.3cm]
\includegraphics[clip=true,bb=120 200 500 660,width=.85\columnwidth,angle=270]{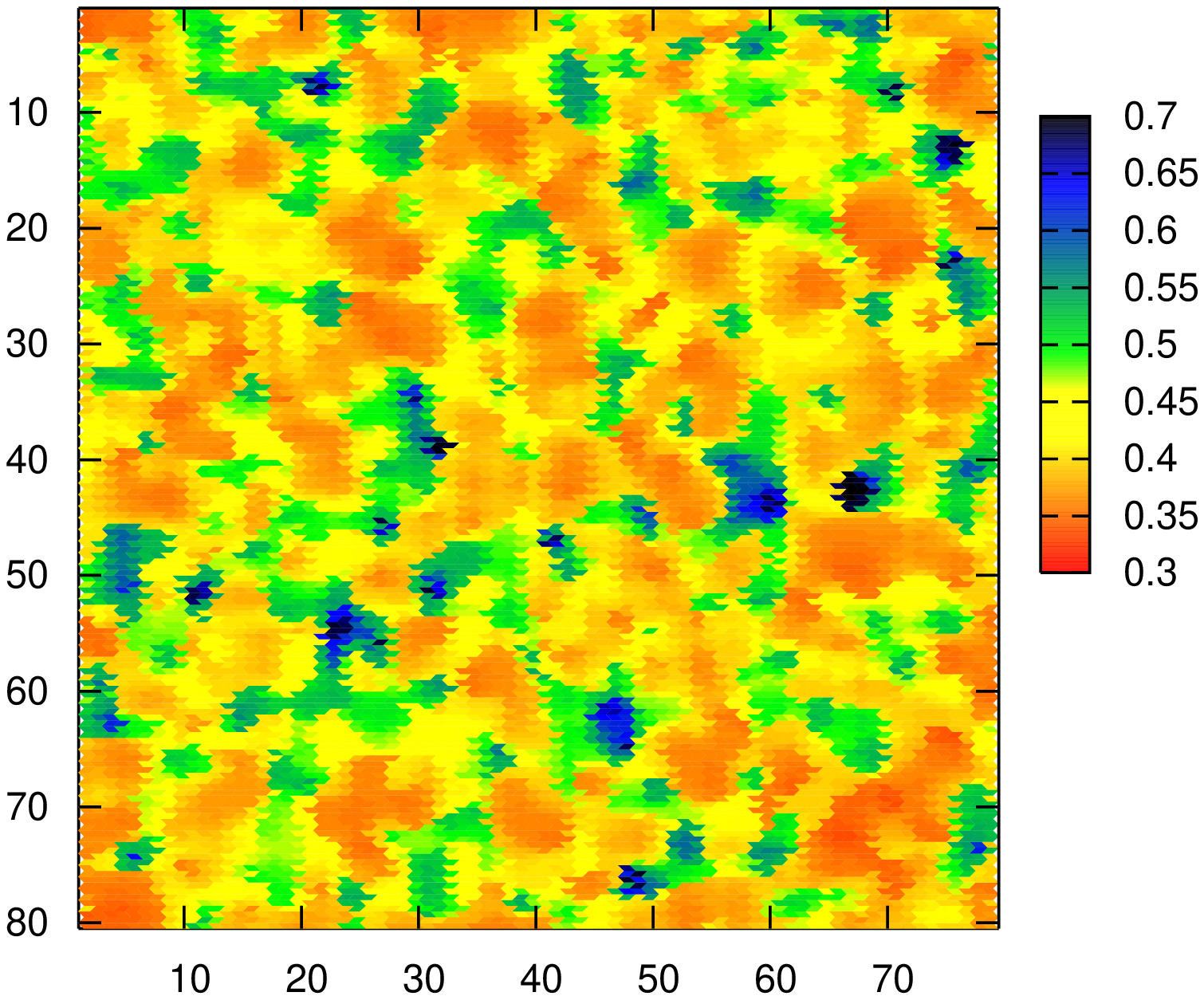}
\end{minipage}
\begin{minipage}{.49\columnwidth}
(b)\\[-.3cm]
\includegraphics[clip=true,bb=120 200 500 660,width=.85\columnwidth,angle=270]{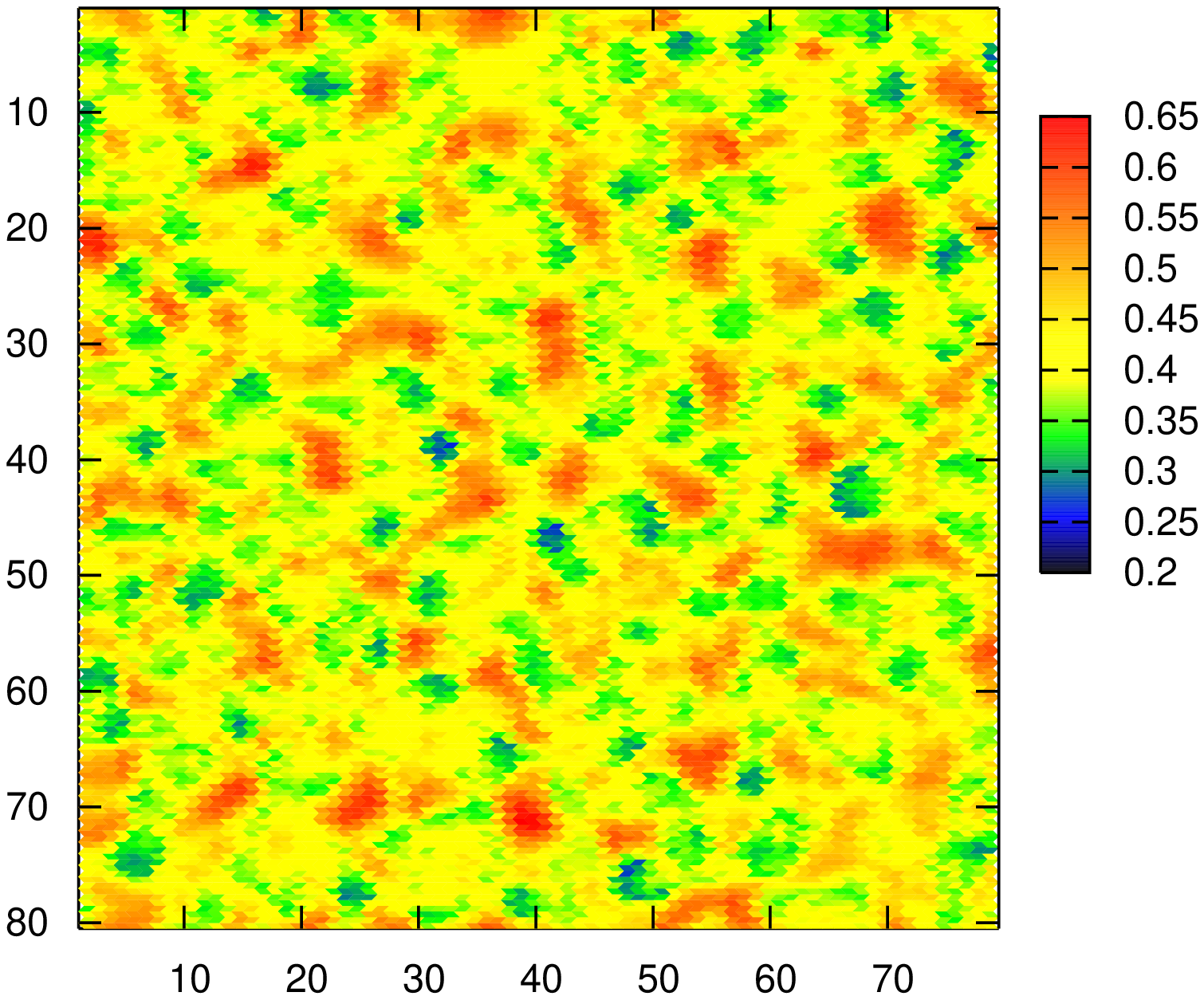}
\end{minipage}\\[-.1cm]
\begin{minipage}{.49\columnwidth}
(c)\\[-.3cm]
\includegraphics[clip=true,bb=120 200 500 660,width=.85\columnwidth,angle=270]{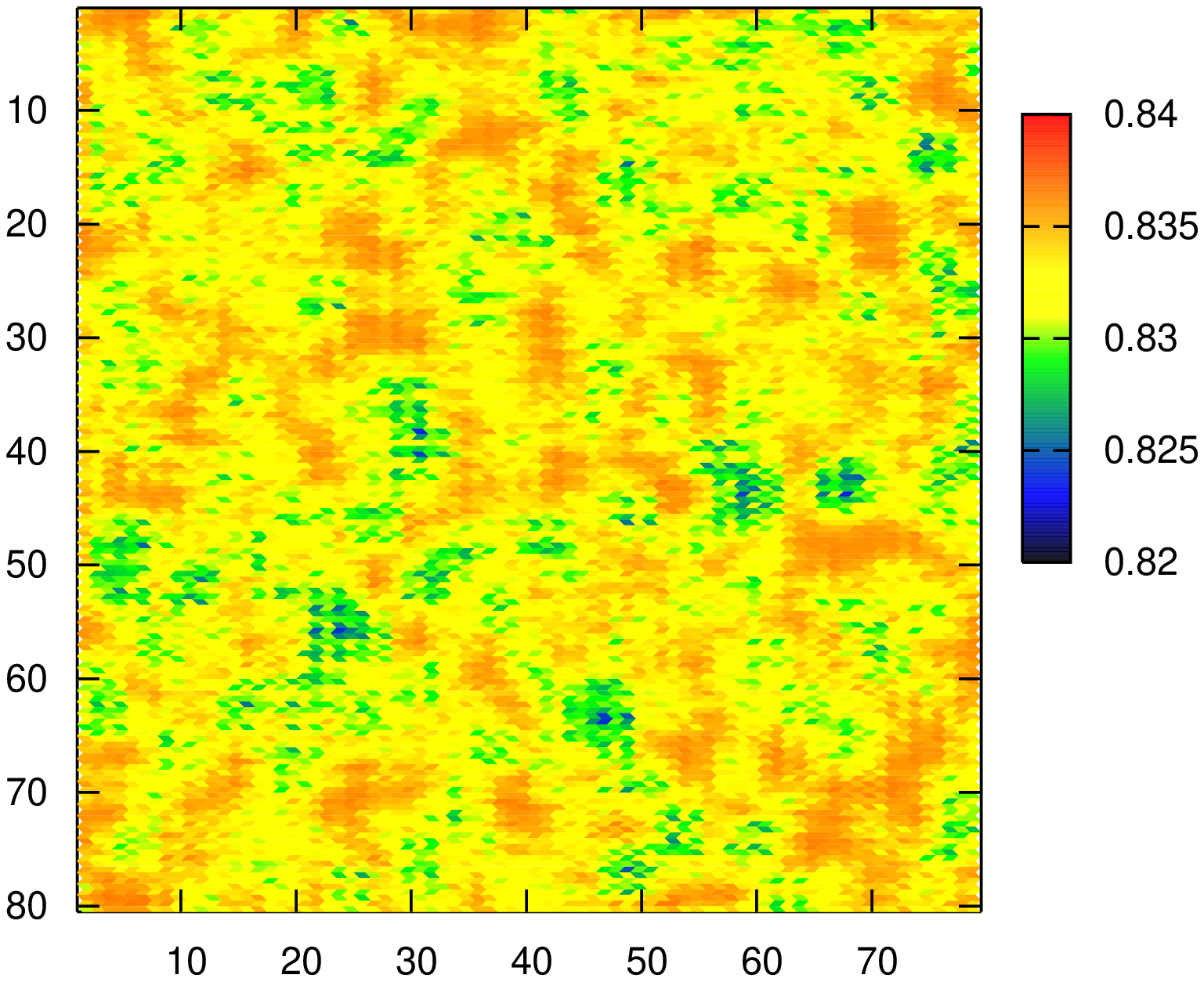}
\end{minipage}
\begin{minipage}{.49\columnwidth}
(d)\\
\includegraphics[clip=true,width=.9\columnwidth]{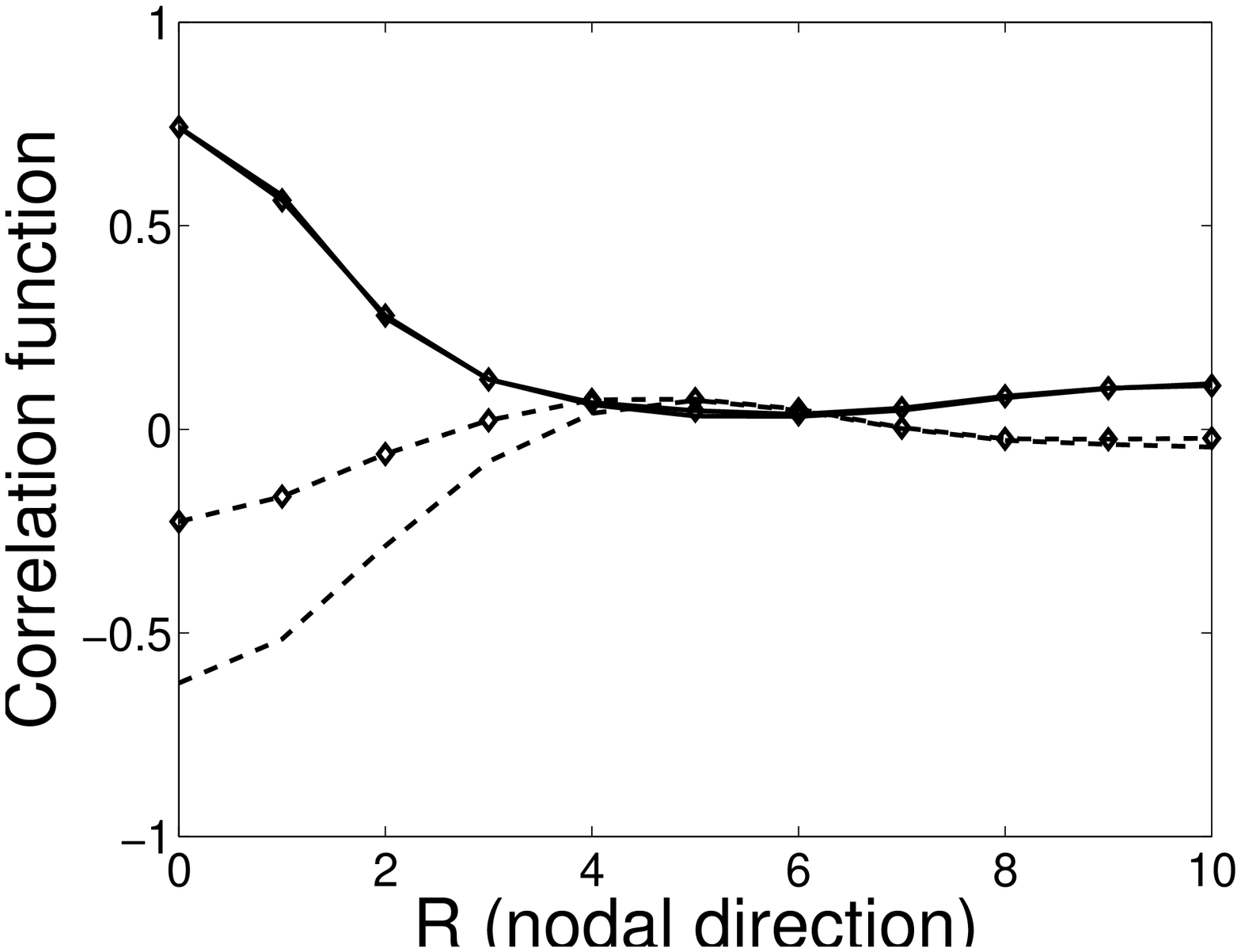}
\end{minipage}
\caption{(Color online.) Many OD impurity model for parameters of
Fig.~\ref{fig:tau3example}(c): (a) 2D real-space
map of the local coherence peak position (gap) in units  of $t$;
(b) coherence peak height (note the inverse color scale with
respect to (a)); (c) total charge (note the small scale); and (d)
the correlation function between the gap map and the dopant atoms
(solid), and the gap map and the peak height map (dashed).  Lines
without (with)  symbols correspond to the parameters  used  in
Fig.~\ref{fig:tau3example}(c) (Fig.~\ref{fig:tau3example}(d)). }
\label{fig:2Dmaps}
\end{figure}

A typical LDOS line scan for a many OD impurity calculation is
shown in Fig.~\ref{fig:tau3example}(c).
Note that, by construction, this model has homogeneous low-energy
LDOS as well as strong correlations between the dopant positions
and the local gap values.
As in the single impurity case, the lineshape of the LDOS near the
gap edge is  determined primarily by Andreev scattering. Because
the LDOS near the gap edge is reminiscent of a coherence peak, we
will simply adopt this terminology, as used in experiment. The
dopant atoms inevitably give rise to a conventional
potential as well, however;  we therefore show in
Fig.~\ref{fig:tau3example}(d) that the qualitative features of OD 
scattering depicted in Fig.~\ref{fig:tau3example}(c) survive this
scattering.
Comparing Figs.~\ref{fig:tau3example}(a-d), it is evident that the
OD LDOS spectra are far more particle-hole symmetric than those
with potential disorder, and display the inverse relation between
gap magnitude and coherence peak height, as expected from the
single-impurity discussion (see Fig.~\ref{fig:TMat}). In
Fig.~\ref{fig:2Dmaps} we show the associated gap map (a), the
coherence peak height map (b), and the charge modulation map (c)
for parameters corresponding to Fig.~\ref{fig:tau3example}(c).
Fig.~\ref{fig:2Dmaps}(d) displays the correlation functions
between the gap map and the dopants, and the gap map and the peak
height map~\cite{corrdef}. The local pairing modulation shown in
Fig.~\ref{fig:2Dmaps} reproduces qualitatively the correct
negative correlation between the gap amplitude and the coherence
peak height, the positive correlation between dopant atom
locations and large gap values, and the relatively small charge
modulations observed in experiment~\cite{DavisAPS}. In addition,
the spectra exhibit the same remarkable particle-hole symmetric
modulations of the coherence peaks observed in
experiment~\cite{davisinhom2}.  This symmetry should manifest
itself in Fourier transform quasiparticle interference patterns
 as well.

In the OD scattering model, short-distance correlations between
the dopant atoms and the gap size are nearly perfect, as seen in
Fig.~\ref{fig:2Dmaps}(d); indeed, they are considerably stronger
than reported in experiment~\cite{DavisAPS}.
This might be due to the difficulty of  identifying all dopant
positions experimentally,  to the presence of additional cation
disorder in BSCCO~\cite{Eisaki}, or to the finite experimental
resolution of the dopant resonances.
The dopant-gap correlations are quite robust against inclusion of
a conventional scattering component (the two solid curves in
Fig.~\ref{fig:2Dmaps}(d) coincide), but the gap-peak height
correlations are rapidly suppressed, as seen in Fig.~\ref{fig:2Dmaps}(d).

A natural question is the extent to which these correlations are
robust against different choices of parameters. We find that the
local spectral properties in the spiky regime of the OD model are
 insensitive to parameters, provided the amplitude of the gap
modulation $\delta\Delta$ is comparable to or larger than the
splitting of the van Hove and coherence peaks in the pure system.
In that case the weight of the van Hove peak is absorbed into the
coherence peak (Fig.~\ref{fig:TMat}(c)).

While we assert the primacy of the OD channel of scattering for
the modulation of the states near the antinode, we emphasize that
nodal quasiparticles are very weakly scattered by this potential,
and so  microwave and thermal transport are probably only
minimally influenced by the effects discussed
here~\cite{nunnermuwave}.  This further implies that the elastic contribution to the ARPES spectral peaks
near the antinodal and nodal points are determined by completely
different scattering processes.

{\it Conclusions.}  The discovery of nanoscale inhomogeneity in
the local electronic structure of
BSCCO-2212~\cite{cren,davisinhom1,Kapitulnik1,davisinhom2} has
provoked an intense discussion about the origin of this phenomenon
in cuprates and related correlated electron materials.   In this
work, we have offered strong evidence that the inhomogeneity in
the coherence peak position is in fact driven by dopant atoms,
located away from the CuO$_2$ plane, whose primary effect on
one-particle properties is to modulate the local pair interaction.
This ansatz allowed us to reproduce, in model single-impurity and
many-impurity calculations, most of the important correlations
observed in recent STM experiments.

The  calculations reported here have been done entirely
within a BCS framework, and as such cannot be expected to
reproduce certain other correlations, such as the increase of the
average gap with underdoping. Nevertheless, we believe that our
results represent  an important step towards further modeling
which may reveal the microscopic nature of the modulated pair
interaction.

{\it Acknowledgments.}  The authors are grateful to Yu.S. Barash,
W. Chen, J.C. Davis, J. Lee,  K. McElroy, J. Slezak, and L. Zhu
for useful discussions. Work was supported by a Feodor-Lynen
Fellowship from the A. v. Humboldt Foundation (TSN) and ONR grant
N00014-04-0060 (PJH,BMA).

\end{document}